# SAR-U-Net: squeeze-and-excitation block and atrous spatial pyramid pooling based residual U-Net for automatic liver segmentation in Computed Tomography

Jinke Wang[1,2], Peiqing Lv[2], Haiying Wang[2], Changfa Shi[3,*]

[1]*Rongcheng College, Harbin University of Science and Technology, Rongcheng, 264300, China*
[2]*School of Automation, Harbin University of Science and Technology, Harbin, 150080, China*
[3]*Mobile E-business Collaborative Innovation Center of Hunan Province, Hunan University of Technology and Business, Changsha, 410205, China*

## Abstract

***Background and objective***: Liver segmentation is an essential prerequisite for liver cancer diagnosis and surgical planning. Traditionally, liver contour is delineated manually by radiologist in a slice-by-slice fashion. However, this process is time-consuming and prone to errors depending on radiologist's experience. In this paper, a modified U-Net based framework is presented, which leverages techniques from Squeeze-and-Excitation (SE) block, Atrous Spatial Pyramid Pooling (ASPP) and residual learning for accurate and robust liver Computed Tomography (CT) segmentation, and the effectiveness of the proposed method was tested on two public datasets LiTS17 and SLiver07.

***Methods***: A new network architecture called SAR-U-Net was designed, which grounded in the classic U-Net. Firstly, the SE block is introduced to adaptively extract image features after each convolution in the U-Net encoder, while suppressing irrelevant regions, and highlighting features of specific segmentation task; Secondly, ASPP was employed to replace the transition layer and the output layer, and acquire multi-scale image information via different receptive fields. Thirdly, to alleviate the gradient vanishment problem, the traditional convolution block was replaced with the residual block, and thus prompt the network to gain accuracy from considerably increased depth.

***Results***: In the LiTS17 database experiment, five popular metrics were used for evaluation, including *Dice coefficient*, *VOE*, *RVD*, *ASD* and *MSD*. Compared with other closely related 2D-based models, the proposed method achieved the highest accuracy. In addition, in the experiment of the SLiver07 dataset, compared with other closely related models, the proposed method achieved the highest segmentation accuracy except for the *RVD*.

***Conclusion***: An improved U-Net network combining SE, ASPP, and residual techniques is developed for automatic liver segmentation from CT images. This new model enables a great improvement on the accuracy compared to 2D-based models, and its robustness in circumvent challenging problems, such as small liver regions, discontinuous liver regions, and fuzzy liver boundaries, is also well demonstrated and validated.

**1 Introduction**

Liver segmentation is a key prerequisite for cancer detection and treatment. In order to acquire accurate information on the liver volume, position and shape from abdominal CT images, radiologist needs to manually label slices one by one. However, this process is labor-intensive and also subject to the radiologist's experience (**Fig. 1** shows three challenging cases). Therefore, there is an urgent clinical need for automatic and accurate liver segmentation methods.

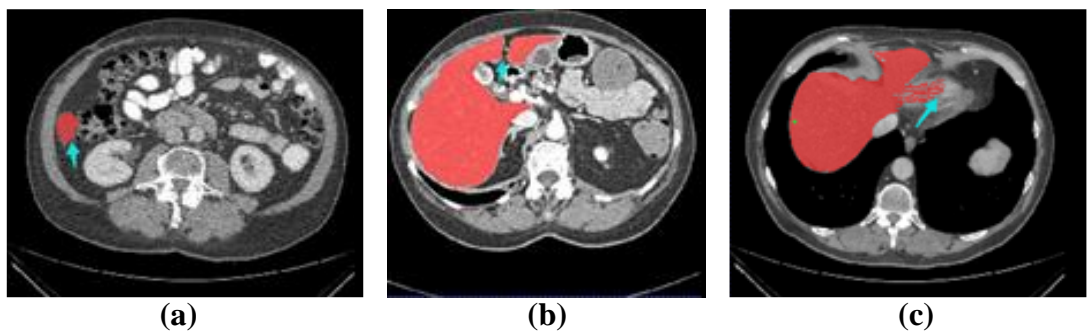

(a)　　　　　　　　　(b)　　　　　　　　　(c)

**Fig. 1.** Three challenging cases, including (a) Small liver area (b) Discontinuity liver area (c) Blurred liver boundary.

In the past few decades, many sophisticated methods have been developed by researchers for automatic liver segmentation, which can be roughly divided into the following three categories: intensity-based methods **[1-5]**, machine learning-based methods **[6-11]**, and deep learning-based methods **[12-22]**.

For intensity-based approaches, they are known for fast speed, including thresholding-based methods **[1-2]**, region growing method **[3]** and level set method **[4-5]**. However, most of them are semi-automatic approaches, which are susceptible to noise, and require human intervention with complicated parameter settings.

For machine learning-based approaches, they enable a great improvement on the segmentation accuracy **[6-11]**. However, most machine learning-based approaches require manual design of specific features, which has a great impact on the accuracy. On the contrary, deep learning-based methods, such as convolutional neural networks (CNN), have developed rapidly due to their intelligent processing of feature engineering, and have achieved a series of successes in tasks such as target detection，image segmentation and classification **[12-14]**.

Among deep learning-based methods, Fully Convolutional Network (FCN) has attracted much attention due to its remarkable pixel-level classification. This model was first proposed by Long et al. **[15]**, and its main difference from CNN is that the fully connected layers of CNN are replaced with the convolutional layers. Ronneberger et al. **[16]** proposed the U-Net model on the basis of FCN, which has achieved great success in the field of medical image segmentation, and has inspired many scholars to participate in the improvement of U-Net model. Milletar et al. **[17]** proposed V-Net, in which residual learning mechanism **[18]** was combined with U-Net to directly achieve end-to-end segmentation of 3D medical images. Christ et al. **[19]** proposed a method of cascading two FCNs to segment the regions of liver and tumor, respectively, and then employed 3D Conditional Random Field (CRF) to optimize the segmentation results. Kaluva et al. **[20]** added dense modules into FCN, and obtained good results in liver and tumors segmentations. Li et al. **[21]** developed the H-Dense U-Net model, which combines 2D U-Net and 3D U-Net models to make full use of the information in and between slices for higher tumor segmentation accuracy, and meanwhile, greatly alleviates the problem of high memory consumption caused by single 3D training method. Han et al. **[22]** designed a 24-layer FCN model, in which skip connections are used between encoder and decoder modules, to achieve the purpose of merging low-level details and high-level semantic information.

Compared with traditional methods, the accuracy of the deep learning-based methods for liver segmentation has been greatly improved. However, there are still limitations for both FCN and U-Net based methods. For the FCN-based method, whether it is single network training or cascade structure training, it's hard to obtain accurate liver results. The main reason is that, while classifying each pixel, FCN does not fully consider the relationship between pixels, which causes the lack of sensitivity to the image details. For the U-Net-based method, although it manages to combine the high-level and low-level semantic information through skip connection, it tends to result in gradient vanishment while blindly stacking network layers, meanwhile, the feature map obtained after U-Net convolution also lacks a specific refining process. In addition, with the continuous down-sampling operation of the network, the resolution of the image gradually decreases, which would adversely affect the segmentation of small regions, and with the increase of the network depth, it is easy to cause the gradient vanishment problem. Finally, the category imbalance problem may also cause errors on small livers regions, discontinuous liver areas and blurred liver boundaries. Although, theoretically, the more z-axis context information is input into 3D network, the better network learning effect, due to the memory constraints, improper selection of the slice number would lead to difficult network training.

The aforementioned methods work well for most automatic liver segmentation situations. However, when it is directly applied to the clinical data, its accuracy and robustness are still insufficient, which heavily hampers the further application of deep learning-based method. To overcome the shortcomings of traditional U-Net in the automatic liver segmentation is exactly what we concern in this paper. Here, we present a novel end-to-end automatic liver segmentation framework, called SAR-U-Net[1], with the main contributions of this paper as follows:

(1) Following the convolution of the encoder in U-Net, the attention mechanism is introduced, so that it can derive image features in an adaptive manner, and meanwhile, suppress irrelevant areas, thus ensure the network could focus on the features related to the specific segmentation task.

(2) Replacing the transition layer and the final output layer of the U-Net decoder with ASPP, so as to achieve the purpose of extracting richer multi-scale feature information.

(3) Replacing the standard convolutional layer of U-Net with the residual block, and attaching a batch normalization layer, which promotes faster convergence, eliminates vanishing gradient problem, and improves segmentation accuracy by training deeper networks.

The manuscript is structured as follows, Section 2 provided the related works on U-Net. Section 3 gives a description of the proposed SAR-U-Net architecture, the loss function, as well as the employed evaluation metrics. The experimental results and discussions are presented in Section 4. Finally, the conclusive remarks and future work are given in Section 5.

## 2 Related Works

We briefly review the most related works, including attention mechanism, atrous convolution and residual learning.

### 2.1 Attention mechanism

In recent years, the framework of attention-based image classification and segmentation have achieved great successes in the field of computer vision **[23-24]**. The attention-based network can not only focus on specific tasks, but also suppress irrelevant areas. According to whether it is differentiable, it could be divided into hard attention and soft attention **[25-26]**. For hard attention, the gradient can be calculated by the neural network, while the weight of attention could be learned via forward propagation and backward feedback **[26]**. Hu et al. **[27]** proposed an attention module named SE block, which automatically obtains the weight of each feature channel, and won the ImageNet2017 competition in image classification task. Noori et al. **[36]** successfully applied the SE Block to the automatic segmentation of brain tumors, as the feature map of the U-Net encoder and the corresponding feature map of the decoder were joined, the SE Block was embedded into it to obtain proper attention weight. Compared with the hard attention mechanism, the soft attention is considered as a global weight calculation approach. Rundo et al. **[37]** proposed the USE-Net, which merged SE into U-Net, and conducted testing with all possible training/testing combinations. They proved that the performance of training on the union of datasets is usually superior to the training on each dataset separately, allowing for both intra-/cross-dataset generalization. Oktay et al. **[28]** developed an attention gate (AG) before splicing the image features obtained by the encoder convolution with the corresponding features in the decoder, and readjusted the output features of the encoder, which was successfully applied to pancreas segmentation in CT. Schlemper et al. **[29]** presented a soft attention mechanism to generate a gated signal that can be trained end-to-end, and effectively validated in fetal ultrasound screening.

---

[1] Our code is publicly available at https://github.com/lvpeiqing/SAR-U-Net-liver-segmentation.

**2.2 Atrous convolution**

The atrous convolution was first proposed by Yu et al. **[32]**, which adds holes (dilations) to the standard convolution to increase the receptive field. It can effectively alleviate the problem of reduced spatial resolution caused by down-sampling. Therefore, atrous convolution is widely used in the field of image segmentation. Chen et al. **[34]** applied the atrous convolution to the deep CNNs model and provided the novel DeepLab V1 model, to obtain more context information. Chen et al. **[30]** made further improvements to the dilated convolution and proposed the ASPP module. For a given input, ASPP can sample the atrous convolution with different sampling rates, which is equivalent to capturing the context of the image at multiple scales. It enables the obtained feature map to contain richer semantic information, which is also crucial for accuracy.

**2.3 Residual learning**

The residual structure was first proposed by He et al. **[18]**. This model could solve the degradation problem while the network depth increases. Jin et al. **[34]** provided a 3D hybrid residual attention-aware segmentation method, which embed residual structure into the basic U-Net, and confirmed a success in liver and tumors segmentations. Bi et al. **[35]** suggested a deep residual network (ResNet) for effective liver lesions segmentation, and ranked the 4th in the LiTS2017 liver segmentation challenge.

In this paper, we leverage the abovementioned three sophisticated techniques, and proposed the SAR-U-Net method for automatic liver segmentation in CT.

**3 Method**
**3.1 SAR-U-Net Architecture**

The proposed network SAR-U-Net is composed of two parts forming a symmetrical structure: the encoder and the decoder part (**Fig. 2**). The encoder is responsible for feature extraction, while the decoder is responsible for feature positioning. The whole architecture consists of 8 residual blocks, 4 pooling layers, 4 SE (Squeeze and Excitation Block), 2 ASPP and multiple up-sampling blocks. The size of convolution kernel is $3\times3$, the pooling layer size is $2\times2$, and the input image is $512\times512\times1$. As each feature image passed through a series of operations such as convolution, feature extraction and pooling, and then the binary segmented image of $512\times512$ is obtained.

In the entire network, the ordinary unit of the traditional U-Net is replaced by the residual learning structure **[18]**. Comparatively, the residual structure adds shortcut connections on the basis of a single forward propagation, so that a deeper network can be trained without degradation, while extracting more discriminative features. In the residual unit, batch normalization and ReLU activation operations are performed following each convolution. By introducing batch normalization, it can not only reduce the sensitivity of the model to the initialization parameters, but also exerts the regularization effect to a certain extent. For the ReLU function, it is most widely used for activation thanks to its ability of circumvent the gradient vanishing problem.

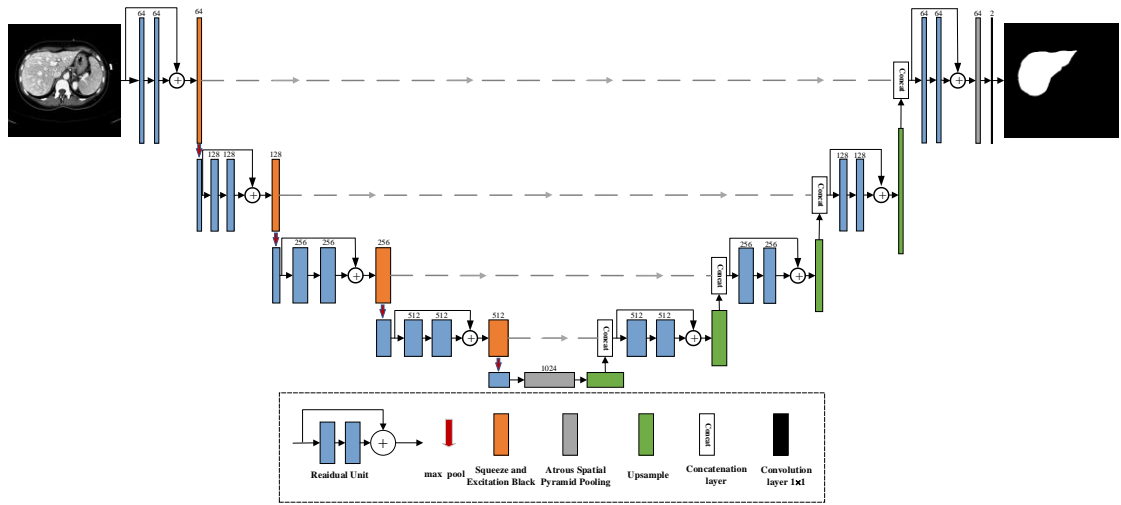

**Fig. 2.** Architecture of proposed SAR-U-Net

In addition, to derive information from the feature map of encoder convolution, SE block **[27]** is utilized after each residual unit, to adaptively extract the image features. Thus, thanks to the channel attention mechanism, the network is enabled to focus on the specific segmentation task (as shown in **Fig. 3**).

The specific operation of the SE module is as follows: First, the 2D feature (H×W) of each channel is compressed into a real number by global average pooling, and then, using a fully connected neural network, a nonlinear transformation is performed to obtain the weight of each feature channel, and finally, the normalized weights obtained above are weighted to the features of each channel, so as to achieve the purpose of extracting specific information.

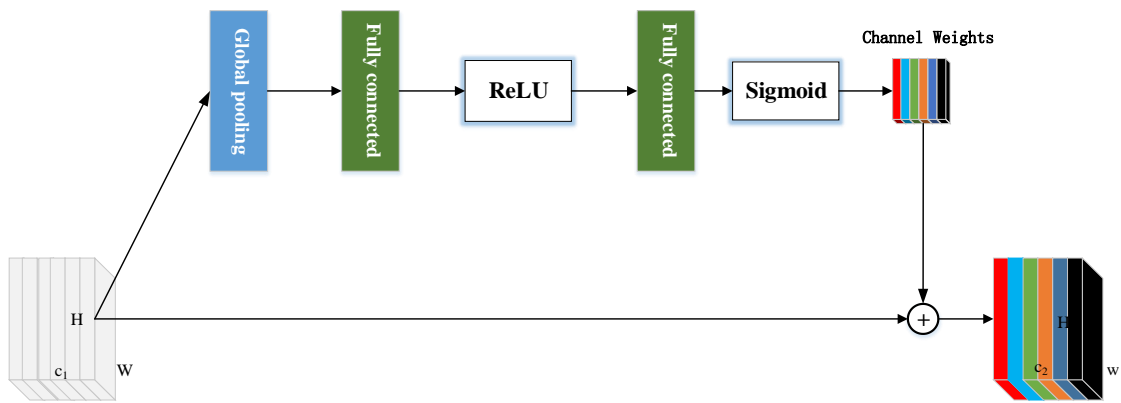

**Fig. 3.** Squeeze and Excitation Block

Furthermore, to alleviate the problem of resolution decrease caused by multiple down-sampling, ASPP **[30]** is utilized as the transition layer of the network (as shown in **Fig. 4**). The ASPP module is able to capture the contextual information of the image at multiple scales, which promotes the inclusion of multi-scale semantic information in the extracted feature map. Similarly, the ASPP module is also introduced in the output of the decoder, by which way, the accuracy is increased together with the transition layer.

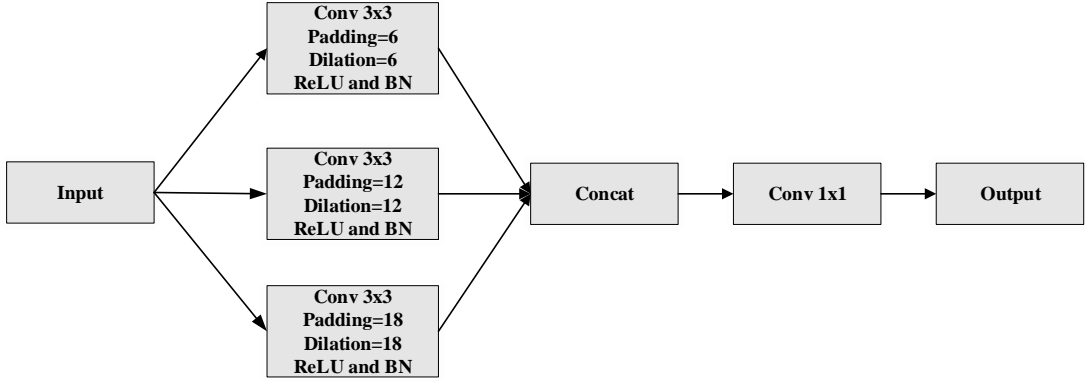

**Fig. 4.** Atrous Spatial Pyramid Pooling

The detailed structure of the proposed network is listed in **Table 1**. The 2nd and 5th columns indicate the size and channels of the output. For the 3rd and 6th columns, [] denotes the convolution operation, while "3×3,64" denotes the convolution kernel of size 3×3 and convolution layer of 64 channels. In addition, the "+" in "Upsample_1+SE_Black4" represents a connection between Upsample_1 and SE_Black4, which results in Concatenate1. Moreover, The "→" in the 3rd column denotes the operation sequence.

**Table 1** SAR-U-Net Network Constructure

|  | Feature size | 2D SAR-U-Net |  | Feature size | 2D SAR-U-Net |
|---|---|---|---|---|---|
| **Input** | 512×512×1 |  | **Upsample_1** | 64×64×1024 | 2×2 |
| **Conv1_x** | 512×512×64<br>512×512×64 | $\begin{bmatrix} 3\times3,64,\text{BN-ReLU} \\ 3\times3,64,\text{BN-ReLU} \end{bmatrix}$ | **Concatenat1** | 64×64×1536 | Upsampe_1+SE_Block4 |
| **SE_Block1** | 512×512×64 | Avg_pool(1,1)→Linear(64,4)<br>→ReLU→Linear(4,16) →<br>Sigmoid | **Conv5_x** | 64×64×512 | $\begin{bmatrix} 3\times3,512,\text{BN-ReLU} \\ 3\times3,512,\text{BN-ReLU} \end{bmatrix}$ |
| **Maxpooling** | 256×256×64 | 2 × 2 | **Upsample_2** | 128×128×512 | 2×2 |
| **Conv2_x** | 256×256×128<br>256×256×128 | $\begin{bmatrix} 3\times3,128,\text{BN-ReLU} \\ 3\times3,128,\text{BN-ReLU} \end{bmatrix}$ | **Concatenat2** | 128×128×756 | Upsample_2+<br>SE_Block3 |
| **SE_Block2** | 256×256×128 | Avg_pool(1,1)→Linear(128,8)<br>→ReLU→Linear(8,128) →<br>Sigmoid | **Conv6_x** | 128×128×256 | $\begin{bmatrix} 3\times3,256,\text{BN-ReLU} \\ 3\times3,256,\text{BN-ReLU} \end{bmatrix}$ |
| **Maxpooling** | 128×128×128 | 2×2 | **Upsample_3** | 256×256×256 | 2×2 |
| **Conv3_x** | 128×128×256<br>128×128×256 | $\begin{bmatrix} 3 \times 3,256,\text{BN-ReLU} \\ 3 \times 3,256,\text{BN-ReLU} \end{bmatrix}$ | **Concatenat3** | 256×256×384 | Upsample_3+<br>SE_Block2 |
| **SE_Block3** | 128×128×256 | Avg_pool(1,1)→Linear(256,16)<br>→ReLU→Linear(16,256) →<br>Sigmoid | **Conv7_x** | 256×256×128 | $\begin{bmatrix} 3 \times 3,128,\text{BN-ReLU} \\ 3 \times 3,128,\text{BN-ReLU} \end{bmatrix}$ |
| **Maxpooling** | 64×64×256 | 2×2 | **Upsample_4** | 512×512×128 | 2×2 |
| **Conv4_x** | 64×64×512<br>64×64×512 | $\begin{bmatrix} 3 \times 3,512,\text{BN-ReLU} \\ 3 \times 3,512,\text{BN-ReLU} \end{bmatrix}$ | **Concatenat4** | 512×512×192 | Upsample_4+SE_Block1 |
| **SE_Block4** | 64×64×512 | Avg_pool(1,1)→Linear(512,32)<br>→ReLU→Linear(32,612) →<br>Sigmoid | **Conv8_x** | 512×512×64 | $\begin{bmatrix} 3 \times 3,64,\text{BN-ReLU} \\ 3 \times 3,64,\text{BN-ReLU} \end{bmatrix}$ |
| **Maxpooling** | 32×32×512 | 2×2 | **ASPP2** | 512×512×64 | |
| **ASPP1** | 32×32×1024 | | **Conv9_x** | 512×512×2 | $[1 \times 1,1]$ |

## 3.2 Loss function

Cross entropy is widely used in various model training by measuring the similarity between the predicted distribution and the ground truth, with its formula as follows,

$$L = -\frac{1}{N}\sum_{i=1}^{N}[\hat{P}_i \log P_i + (1-\hat{P}_i)\log(1-P_i)] \quad (1)$$

where $P_i$ represents the probability of voxel $i$ belongs to the foreground, while $\hat{P}_i$ represents the gold standard, and $N$ denotes the number of samples.

However, in the actual liver segmentation, due to the unbalanced ratio of foreground (liver) and background (non-liver), it is likely to cause low accuracy in the segmentation of small liver regions. Therefore, the following balancing operations are performed, in which an additional weight factor $\omega_i^{class}$ [38] is introduced to weight the loss function in the original **Eq. (1)**, with the formular as follows,

$$L = -\frac{1}{N}\sum_{i=1}^{N}\omega_i^{class}[\hat{P}_i \log P_i + (1-\hat{P}_i)\log(1-P_i)] \quad (2)$$

$$w_i^{class} = \frac{N-n_i}{n_i} \quad (3)$$

where $n_i$ represents the total number of pixels belonging to class $i$, while $N$ represents the total number of pixels belonging to all classes. Specifically, $\omega_i$ is the ratio of the total pixel number outside the category ($N-n_i$) to the total pixel number of all categories ($n_i$). Therefore, the more pixels of category $i$, the smaller the weight coefficient $\omega_i$. Since the number of background pixels is much larger than that of liver pixels in liver CT, the weight factor $\omega_i$ can effectively overcome the imbalance between background and target, and thus improve the segmentation accuracy.

**3.3 Evaluation Metrics**

For liver segmentation evaluation, five most commonly used metrics are adopted, including Dice coefficient (*DC*), Volume Overlap Error (*VOE*), Relative Volume Error (*RVD*), Average Symmetric Surface Distance (*ASD*) and Maximum Surface Distance (*MSD*). Assuming that, *A* is the segmentation result of the liver, and *B* is the ground truth, then the definitions of the five metrics are as follows:

(1) *Dice coefficient* （*DC*）: the similarity of two sets, whose range is [0,1]. The larger the value, the higher the segmentation accuracy.

$$Dice(A, B) = \frac{2|A \cap B|}{|A| + |B|} \tag{4}$$

(2) *Volume Overlap Error (VOE)*: the difference between the segmented volume and the ground truth volume.

$$VOE(A, B) = 1 - \frac{|A \cap B|}{|A \cup B|} \tag{5}$$

(3) *Relative Volume Error (RVD)*: Used to determine whether the segmentation result is over-segmented or under-segmented. The closer the value is to zero, the higher accuracy of the segmentation.

$$RVD(A, B) = \frac{|B| - |A|}{|A|} \tag{6}$$

(4) *Average Symmetric Surface Distance (ASD)*: the average distance between the surfaces of segmentation results *A* and the ground truth *B*, where *d* (*v*, *S(X)*) represents the shortest Euler distance from voxel *v* to the surface voxel of the segmentation result.

$$ASD(A, B) = \frac{1}{|S(A)| + |S(B)|} \left( \sum_{p \in S(A)} d(p, S(B)) + \sum_{q \in S(B)} d(q, S(A)) \right) \tag{7}$$

(5) *Maximum Surface Distance (MSD)*: the maximum surface distance between the segmentation results *A* and the ground truth *B*.

$$MSD(A, B) = \max \left\{ \max_{p \in S(A)} d(p, S(B)), \max_{q \in S(B)} d(p, S(A)) \right\} \tag{8}$$

**4. Experiment**

**4.1 Dataset and implementation**

In the experiment, the labeled training sets of the LiTS17 [2] and SLiver07[3] datasets are used. The LiTS17-Training dataset consists of 131 sets of 3D abdominal CT scans, with a variety of different sampling protocols. The size of each CT image and labels is 512×512, and the in-plane resolution is 0.55mm ~1.0mm, with the inter-slice spacing 0.45mm ~ 6.0mm. In our experiment, 121 datasets were randomly selected from the 131 training datasets for training, while the other 10 datasets are used for testing. In addition, the SLiver07-Training dataset consists of 20 sets of datasets, in which the size of each CT image is 512×512, and the in-plane resolution is 0.56mm ~ 0.8mm, with the inter-slice spacing 1mm ~ 3mm, and all the 20 datasets were used for testing.

---
[2] https://competitions.codalab.org/competitions/17094#results
[3] http://www.sliver07.org/

In the training phase, the initial value of the learning rate (*lr*) is set to 0.001, and is attenuated according to the formula $lr = initial\_lr \times \gamma^{epoch/step\_size}$, where the initial value (*initial_lr*) is set to 0.5, and the *step_size* is set to 4. We use standard SGD (Stochastic Gradient Descent) to optimize the objective function. The *batch size* is empirically set to 4, which is the maximum batch size that the proposed method can run on our GPU with 11GB memory. Since most trainings converge around epoch of 60, we empirically set *epoch* to 60 to ensure effective training. All the experiments are run on a workstation with Ubuntu 18.04 operating system, graphics card RTX2080Ti, memory 64G, single CPU Intel Xeon Silver 4110, and using the Pytorch1.4 deep learning framework for implementation.

## 4.2 Image preprocessing

In the preprocessing stage, first, to remove irrelevant organs, the Hounsfield intensity is transformed to [-200, 200], and histogram equalization is applied to enhance the contrast and clarity of the image (as shown in **Fig. 5(b)**). Then, the z-axis spacing of all data is resampled to 1mm, and the intensity of the image is normalized to [0,1] (as shown in **Fig. 5(c)**). Finally, in order to further optimize the model weight during training, all the pixel values are standardized.

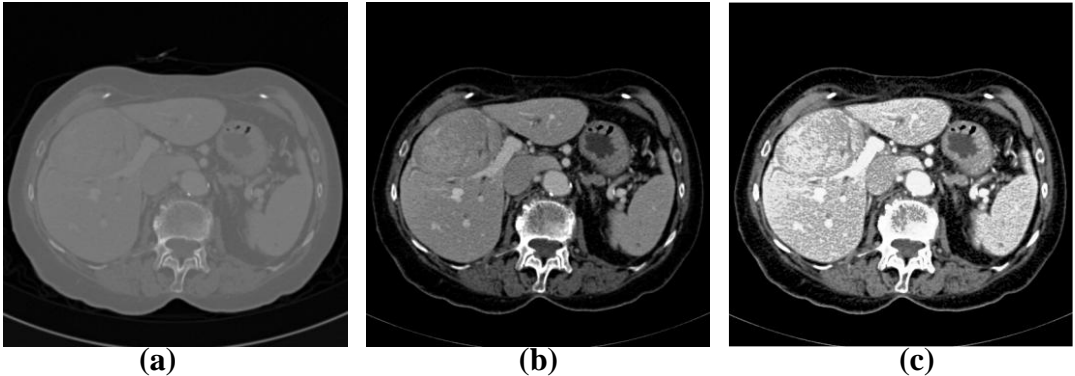

           **(a)**                        **(b)**                       **(c)**

**Fig. 5**. Flowchart of preprocessing. (a) Original CT (b) HU window processing (c) Histogram equalization

## 4.3 Dataset augmentation

To eliminate the overfitting problem in the training process, data augment is an alternative strategy via expanding the training datasets. In this step, a variety of rigid and elastic transformations were utilized, including: (1) scaling the image between 0.8 and 1.2 with a 50% probability (**Fig. 6(b-c)**), (2) rotating the image between 0 degrees and 30 degrees with a 30% probability (**Fig. 6(d)**), (3) turning the image left-right or up-down with a 30% probability (**Fig. 6(e)**), and (4) B-spline elastic deformation **[31]** (**Fig. 6(f)**).

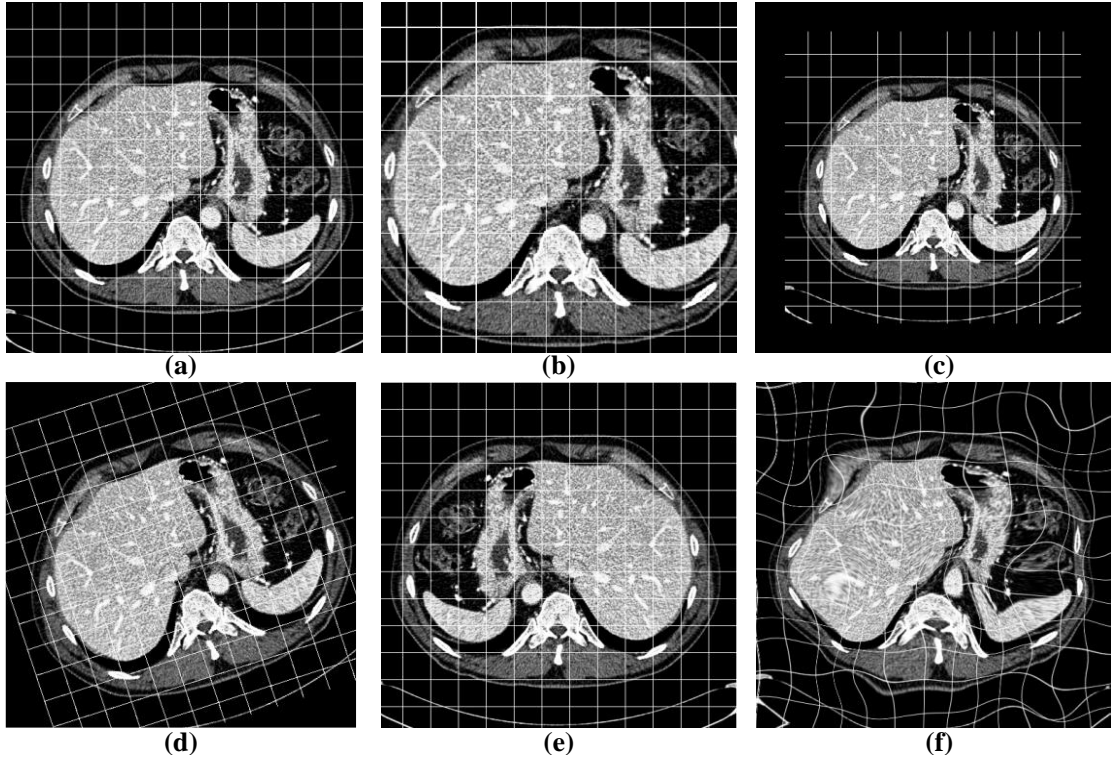

**Fig. 6.** Data augment. (a) Original CT (b) Zoom in (c) Zoom out (d) Rotation (e) Rolling over (f) Elastic deformation

**4.4 Test on LiTS17-Training dataset**

The proposed method is compared with three related classic models, including the FCN[4] model based on the VGG framework **[15]**, U-Net[5] **[16]** and Attention U-Net[6] **[28]**. To ensure a fair comparison and reproductivity of the experiment, the open-source codes of the above models were provided and used for training and testing with the same datasets, respectively. The input of all training models is 2D image, while the lesion label and the liver label are merged for liver training purpose.

*4.4.1 Quantitative analysis of segmentation accuracy*

**Table 2** shows the quantitative comparison results of the four models. It can be seen from **Table 2** that, our proposed model showed higher performances in the comparisons on five evaluation metrics, especially for the *Dice*, the proposed method obtains the highest mean value of 95.71. Compared with the other three models, the FCN model shows the lowest performances in *Dice*, *VOE*, *RVD* and *MSD*. Compared with U-Net and Attention U-Net, the *Dice* (*MSD*) of our proposed model is much higher (lower), while the *VOE*, *ASD*, and the absolute value of *RVD* are also slightly lower. A detailed results of the proposed model using the LiTS17 datasets is also provided in **Table 3**. The box plot is also provided in **Fig. 7.** It can be seen from the figure that, compared with the other three models, the performance of the proposed model on all five-evaluation metrics have significantly improved with small variance, which proves the strong robustness of SAR-U-NET.

---

[4] https://github.com/shelhamer/fcn.berkeleyvision.org
[5] https://github.com/milesial/Pytorch-U-Net
[6] https://github.com/ozan-oktay/Attention-Gated-Networks

**Fig. 8 (a)(b)** and **Fig. 8 (c)(d)** depict the loss curve and accuracy curve of different models on the training set and validation set, respectively. It can be seen from the figures that the loss value of the proposed SAR-U-NET shows smoother curve with higher accuracy, and the epoch tends to be stable at about 20, proving that SAR-U-NET is more robust and more accurate than FCN, U-Net and Attention U-Net.

In the comparative experiment, the statistical method of *t*-test is also used to verify whether the proposed method is significantly different from other methods on the accuracy. All statistical hypothesis tests are based on the representative metrics of *Dice* and *ASD*. As shown in **Table 2**, the performance difference between the labeled results and our proposed results is statistically significant ($p < 0.05$).

Table 2 Quantitative results among the four methods on 10 LiTS17-Training datasets

| Methods | Dice (%) | VOE (%) | RVD (%) | ASD (mm) | MSD (mm) |
|---|---|---|---|---|---|
| FCN[15] | 88.63±4.28* | 20.15±6.84 | -15.57±9.28 | 3.46±1.05* | 38.23±3.29 |
| U-Net[16] | 92.60±2.54* | 10.33±1.69 | -0.28±6.13 | 2.41±0.54* | 33.69±1.89 |
| Attention U-Net[28] | 94.71±0.72* | 10.25±1.30 | -1.69±3.13 | 2.45±0.46* | 30.25±2.62 |
| **SAR-U-NET** | **95.71±0.55** | **9.52±1.11** | **-0.84±3.86** | **1.54±0.30** | **29.14±2.63** |

Results are represented as mean and standard deviation. Note: ∗ indicates a statistically significant difference between the marked result and the corresponding one of our method at a significance level of 0:05.

Table 3 The results of our proposed method on 10 LiTS17-Training datasets

| Case Num | Dice | VOE | RVD | ASD (mm) | MSD (mm) |
|---|---|---|---|---|---|
| 1 | 0.955 | 0.103 | -0.029 | 1.568 | 28.453 |
| 2 | 0.963 | 0.089 | -0.033 | 1.471 | 34.951 |
| 3 | 0.956 | 0.100 | -0.049 | 1.382 | 31.259 |
| 4 | 0.968 | 0.097 | -0.048 | 1.494 | 25.494 |
| 5 | 0.949 | 0.114 | -0.040 | 1.797 | 28.315 |
| 6 | 0.953 | 0.107 | -0.038 | 1.933 | 29.756 |
| 7 | 0.960 | 0.094 | 0.034 | 1.229 | 30.657 |
| 8 | 0.961 | 0.073 | 0.043 | 0.955 | 39.421 |
| 9 | 0.954 | 0.086 | 0.036 | 1.673 | 34.598 |
| 10 | 0.952 | 0.090 | 0.040 | 1.863 | 28.534 |
| **Avg** | **0.957** | **0.095** | **-0.0084** | **1.544** | **29.144** |

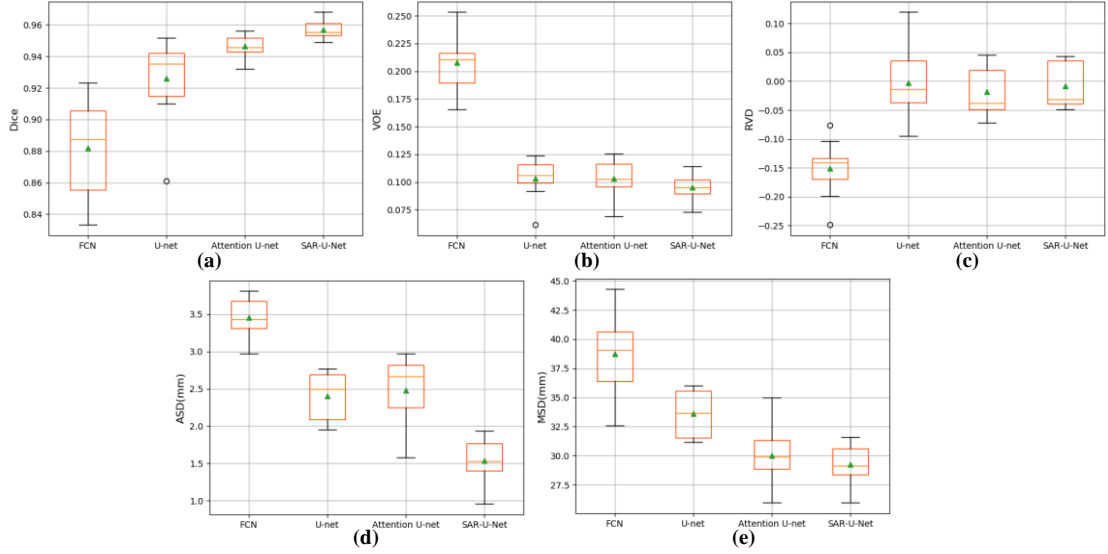

**Fig. 7.** Comparative liver segmentation results (a) Dice (b)VOE (c) RVD (d) ASD (e) MSD

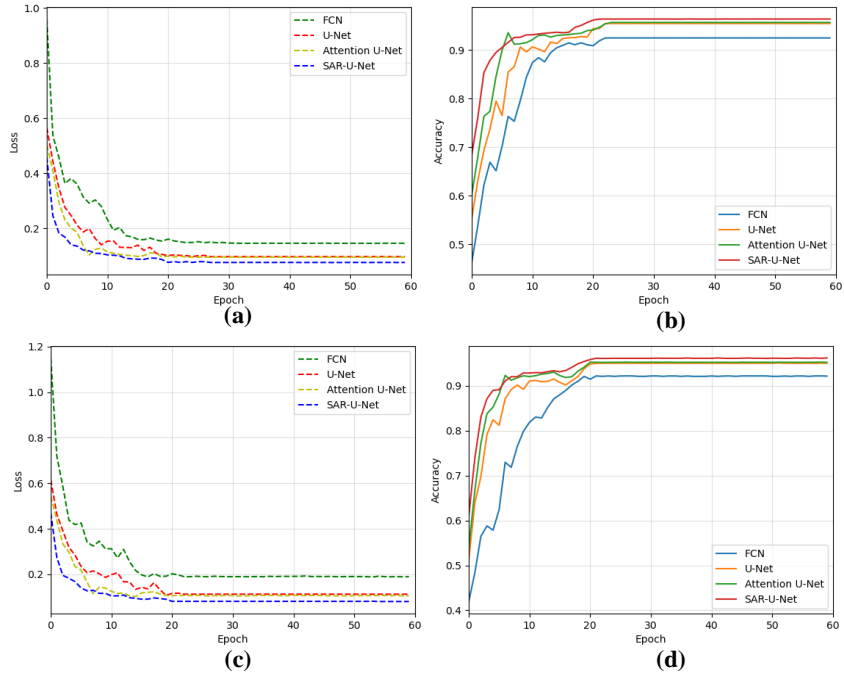

**Fig. 8.** Loss and accuracy curves of different models on LiTS17 datasets (a) loss in training set (b) accuracy in training set (c) loss in validation set (d) accuracy in validation set.

*4.4.2 Visual segmentation results on challenging liver cases*

**Fig. 9** provides visual segmentation comparisons between the proposed method and other three classic algorithms when handling challenging problems. The first and second rows in **Fig. 9** show the comparison results of the small liver area. It can be seen that, all the models are able to segment the liver area properly, except for partial over-segmentation or under-segmentation that occur in FCN, U-Net and Attention U-Net models. In contrast, our proposed model successfully circumvents such errors.

In addition, the third and fourth rows of images in **Fig. 9** show the comparison results of discontinuous regions of the liver. It can be seen that, the FCN model performs the worst. The FCN model not only misses the liver area, but also leads to typical under-segmentation errors. Although U-Net, Attention U-Net and our proposed models shows significant advantages when dealing with organs adjacent to the liver, some under- segmentation still occurred.

Furthermore, the 5th and 6th rows of **Fig. 9** show the comparison results with blurred liver boundaries. It can be seen that, FCN and U-Net perform relatively poorly, with typical over- and under-segmentation errors. However, thanks to the attention mechanism, Attention U-Net showed a slight under-segmentation error, and the proposed method successfully avoided the error.

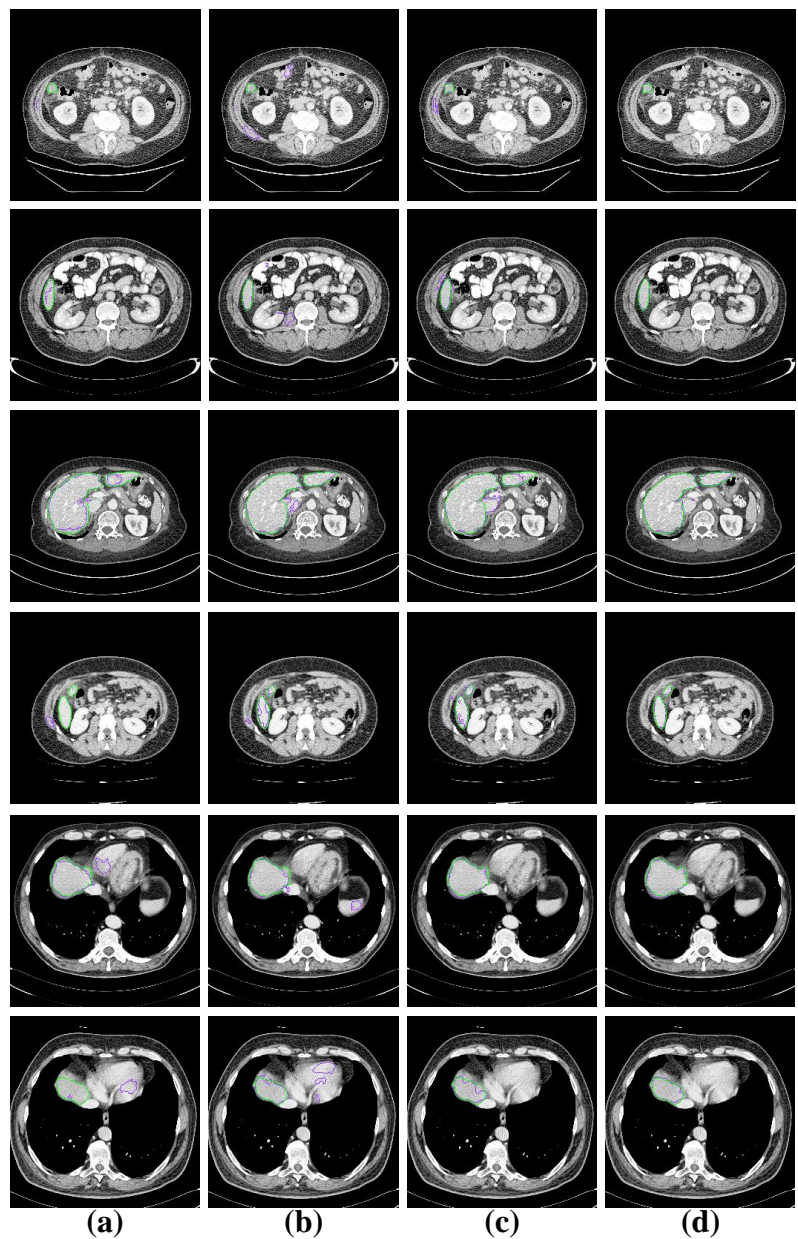

(a)      (b)      (c)      (d)

**Fig. 9.** Comparison of the difficult segmentation problems of the four algorithms. (The first and second rows of images show small liver areas, the third and fourth rows show discontinuous areas of the liver, and the fifth and sixth rows show blurred areas of liver boundaries), where green represents the gold standard, and purple represents the results of different methods. (a) FCN (b) U-Net (c) Attention U-Net (d) SAR-U-Net

Therefore, our proposed model shows relatively high robustness when dealing with difficult segmentation problems such as small liver regions, discontinuous liver areas, and blurred liver boundaries. The main reason is as following, firstly, the attention mechanism is introduced in the encoder, so that the network can pay more attention to the specific features of liver segmentation. Secondly, the atrous spatial convolutional pooling pyramid layer is utilized to extract multi-scale feature information, which effectively compensates for the loss of important combination information. Finally, the addition of the residual structure enables the derivation of more complex image features.

**4.5 Ablation analysis on LiTS17-Training datasets**

To verify the optimality of the proposed network configuration, five comparative ablation experiments were performed. Take U-Net as the baseline and make changes one by one for testing the contribution of the improved module to the overall framework. First, all convolution units in the U-Net encoder and the decoder parts are replaced with residual structures (ResU-Net). Then, the attention mechanism (SE-ResU-Net) is introduced on the basis of the ResU-Net model, which followed the convolution operation of the encoder. At last, on the basis of ResU-Net, the transition layer connecting the encoder and the decoder and the output layer of the decoder are replaced with ASPP modules. All of the models are trained and evaluated on the LiTS17 dataset. From the box plots in **Table 4** and **Fig. 10**, it can be seen that, comparing to a single utilization, the combination of SE and ASPP mechanism significantly improved the performance of the U-Net-based model.

**Fig. 11 (a)(b)** and **Fig. 11 (c)(d)** depict the loss curve and accuracy curve of different models on the training set and validation set, respectively. It can be seen from the figures that, with the improvement of network structure, the curves of loss and accuracy become smoother. Thanks to the use of residual, attention module and ASPP modules, the proposed SAR-U-NET becomes deeper, and more image features are extracted, and thus achieves the lowest loss value and the highest accuracy.

Table 4. Quantitative analysis results of ablation experiments

| Model | Dice (%) | VOE (%) | RVD (%) | ASD (mm) | MSD (mm) |
|---|---|---|---|---|---|
| U-Net | 92.60±2.54* | 10.33±1.69 | -0.28±6.13 | 2.41±0.54* | 33.69±1.89 |
| ResU-Net | 93.85±1.04* | 11.47±2.02 | -0.78±4.88 | 1.94±0.47* | 32.73±1.94 |
| SE-ResU-Net | 93.71±0.98* | 11.81±1.73 | -1.02±4.22 | 1.85±0.45* | 31.05±1.56 |
| ASPP-ResU-Net | 94.20±0.51* | 10.12±1.10 | -1.54±3.68 | 1.67±0.34* | 30.70±2.35 |
| **SAR-U-NET** | **95.71±0.55** | **9.52±1.11** | **-0.84±3.86** | **1.54±0.30** | **29.14±2.63** |

Results are represented as mean and standard deviation. Note: ∗ indicates a statistically significant difference between the marked result and the corresponding one of our method at a significance level of 0:05.

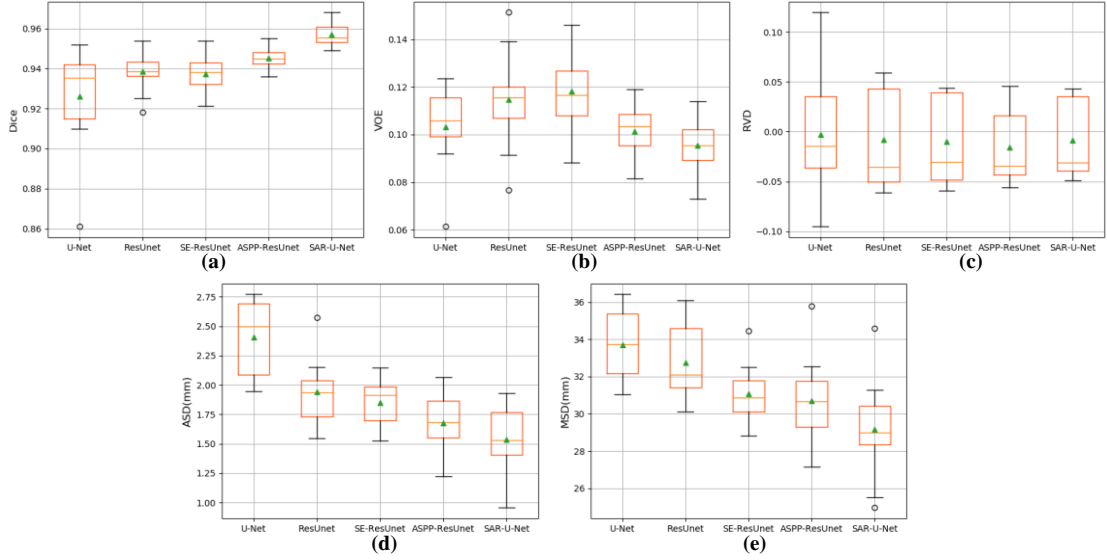

**Fig. 10.** Comparative liver segmentation results (a) Dice (b)VOE (c) RVD (d) ASD (e) MSD

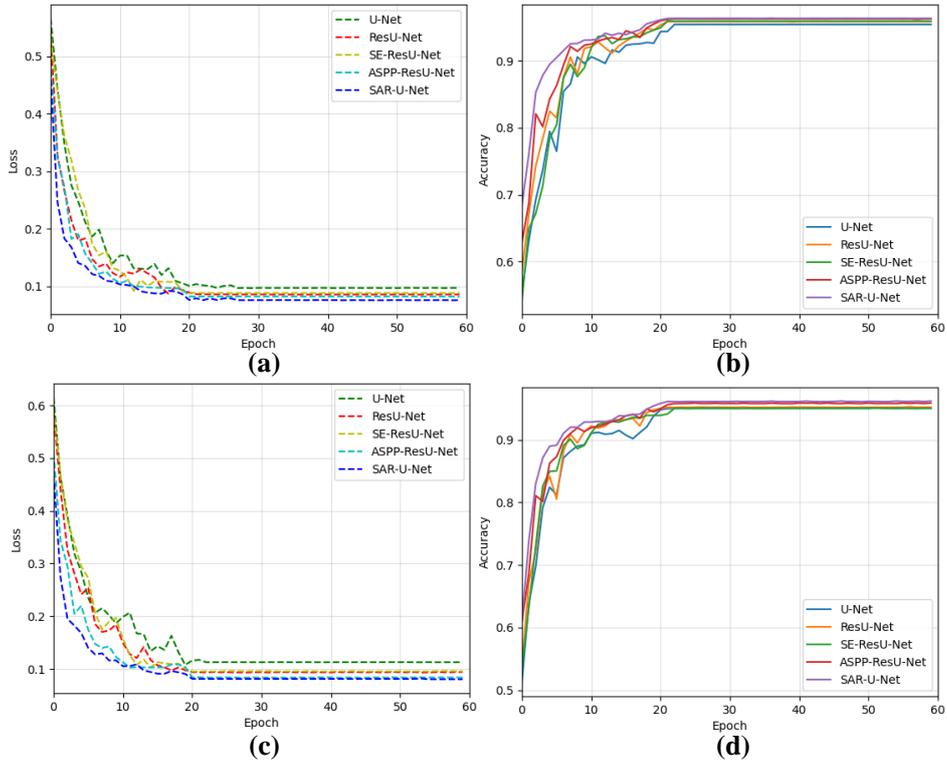

**Fig. 11.** Loss and accuracy curves of different models in ablation analysis of LiTS17 datasets (a) loss in training set (b) accuracy in training set (c) loss in validation set (d) accuracy in validation set.

### 4.6 Test on SLiver07-Training dataset

To further verify the generalization ability of the proposed method, a comparative experiment was conducted using FCN, U-Net, Attention U-Net, and our proposed model on 20 SLiver07-Training datasets. **Table 5** and **Fig. 12** shows the quantitative comparative results among our method and the other three methods. It can be seen from **Table 5** and **Fig. 12** that, compared with other five methods, our proposed method achieved the highest *Dice* of 97.31, which further proves the robustness of our method, with a detailed results of all the metrics shown in **Table 6**.

**Fig. 13 (a)(b)** and **Fig. 13 (c)(d)** depict the loss curve and accuracy curve of different models on the training set and validation set, respectively. It can be seen from the figures that the loss value curve of the proposed model is the smoothest, with the fastest convergence and the highest accuracy, which further proves the effectiveness of the proposed method.

**Table 5.** Quantitative comparison with three state-of-the-arts methods on 20 Sliver07-Training datasets

| Methods | Dice (%) | VOE (%) | RVD (%) | ASD (mm) | MSD (mm) |
|---|---|---|---|---|---|
| FCN[15] | 95.60±3.41* | 8.23±5.89 | -2.38±2.16 | 2.19±0.38 | 36.69±1.45 |
| U-Net[16] | 96.94±1.78* | 5.31±3.48 | -0.54±2.24 | 1.95±0.34 | 30.66±2.03 |
| Attention U-Net[28] | 96.66±2.19* | 6.38±3.93 | -1.29±3.58 | 1.80±0.38 | 28.30±2.05 |
| **SAR-U-Net** | **97.31±1.49** | **5.37±3.27** | **-1.08±2.06** | **1.85±0.30** | **27.45±1.89** |

Results are represented as mean and standard deviation. Note: ∗ indicates a statistically significant difference between the marked result and the corresponding one of our method at a significance level of 0:05.

**Table 6.** Result of our proposed SAR-U-Net on 20 SLiver07-Training datasets

| Test case | Dice | VOE | RVD | ASD (mm) | MSD (mm) |
|---|---|---|---|---|---|
| 1 | 0.945 | 0.103 | -0.059 | 1.783 | 27.935 |
| 2 | 0.952 | 0.091 | -0.067 | 2.284 | 36.534 |
| 3 | 0.975 | 0.048 | -0.024 | 2.138 | 27.365 |
| 4 | 0.978 | 0.042 | -0.001 | 1.538 | 25.934 |
| 5 | 0.934 | 0.156 | 0.011 | 1.959 | 29.446 |
| 6 | 0.977 | 0.043 | -0.001 | 1.798 | 24.551 |
| 7 | 0.979 | 0.040 | -0.013 | 2.139 | 30.316 |
| 8 | 0.974 | 0.050 | -0.023 | 2.502 | 25.233 |
| 9 | 0.982 | 0.035 | -0.011 | 1.618 | 27.428 |
| 10 | 0.946 | 0.101 | 0.010 | 2.058 | 26.452 |
| 11 | 0.985 | 0.029 | -0.005 | 1.973 | 31.563 |
| 12 | 0.984 | 0.031 | 0.009 | 1.470 | 26.521 |
| 13 | 0.972 | 0.054 | 0.001 | 1.387 | 27.078 |
| 14 | 0.973 | 0.051 | -0.020 | 2.059 | 28.726 |
| 15 | 0.985 | 0.028 | 0.005 | 2.107 | 25.459 |
| 16 | 0.984 | 0.031 | -0.003 | 1.474 | 24.234 |
| 17 | 0.980 | 0.037 | 0.001 | 1.841 | 29.643 |
| 18 | 0.981 | 0.036 | -0.003 | 1.759 | 28.694 |
| 19 | 0.988 | 0.024 | 0.003 | 1.636 | 28.419 |
| 20 | 0.979 | 0.041 | -0.025 | 1.469 | 27.486 |
| Avg | 0.973 | 0.054 | -0.0108 | 1.849 | 27.449 |

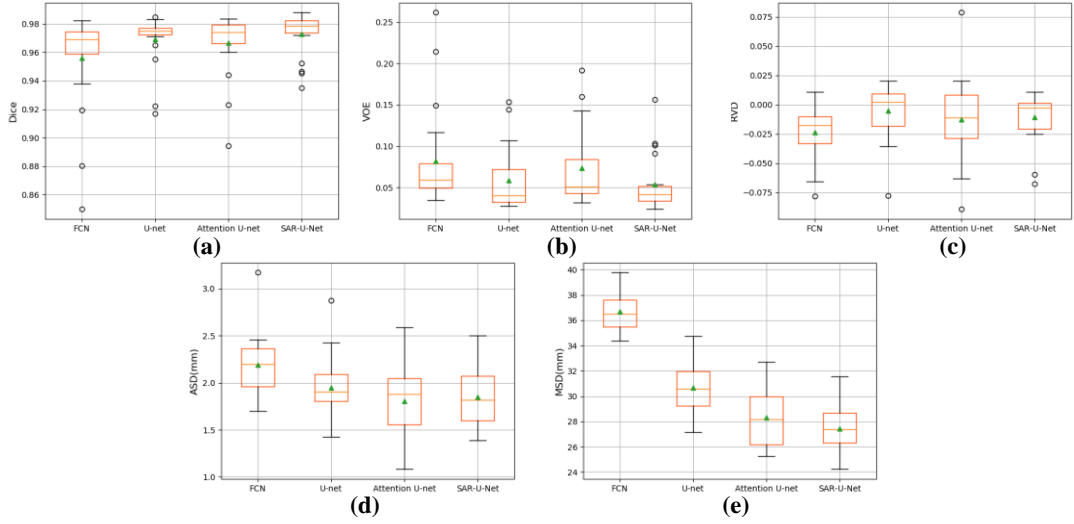

**Fig. 12.** Comparative liver segmentation results. (a) Dice (b)VOE (c) RVD (d) ASD (e) MSD

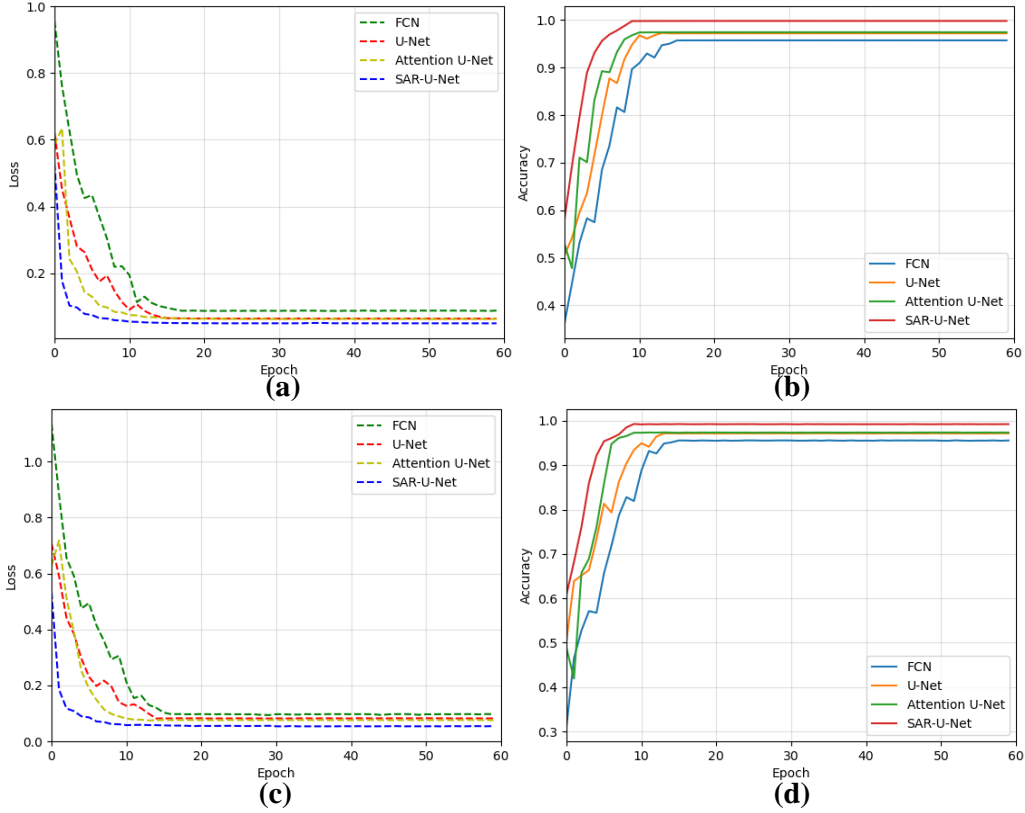

**Fig. 13.** Loss and accuracy curves of different models in ablation analysis of LiTS17 datasets (a) loss in training set (b) accuracy in training set (c) loss in validation set (d) accuracy in validation set.

**Fig. 14** shows the comparative results of 2D visualization among FCN, U-Net, Attention U-Net, and our proposed method. It can be seen that, for small liver regions, discontinuous liver regions and blurred liver boundaries, our proposed method shows high robustness in handling these challenging cases.

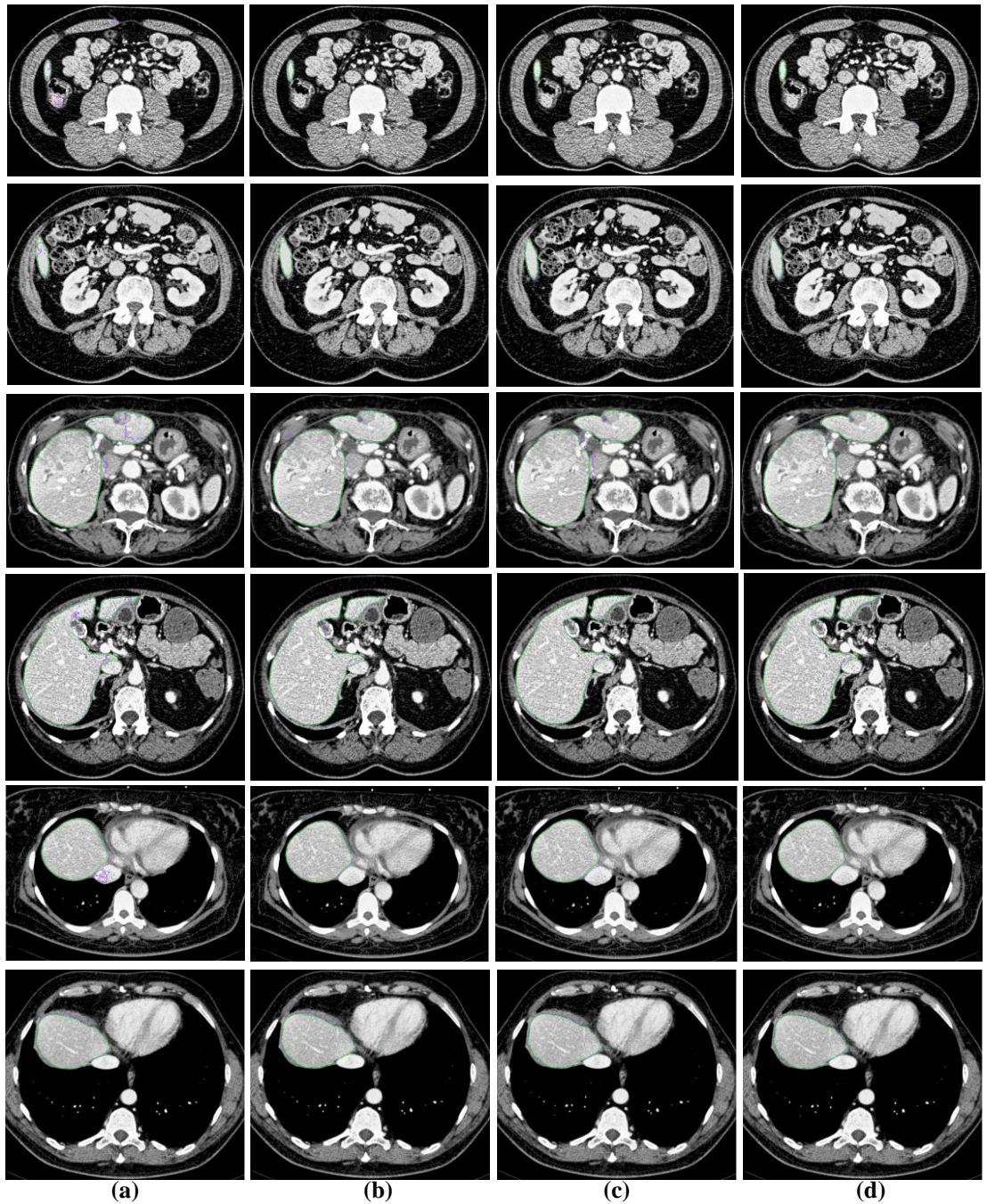

(a)    (b)    (c)    (d)

**Fig. 14**. Test on difficult cases of SLiver07. The first and second rows denote the small liver area, the third and fourth rows denote the discontinuous liver area, and the fifth and sixth rows denote the blurred liver boundary, where the green line represents the gold standard, and the purple line represents the result of our proposed method. (a) FCN (b) U-Net (c)Attention U-Net (d) **SAR-U-NET**

### 4.7 Impact of weight factor on the loss function

To prove the improvements the weight factor bring to the system performance, comparative experiments based on SAR-U-NET using Binary Cross-Entropy (BCE) and Weighted Binary Cross-Entropy (WBCE) were implemented. **Fig. 15(a)(b)** and **Fig. 15(c)(d)** showed the loss value and accuracy on the training and validation sets in LiTS17 and SLiver07, respectively.

From the loss function curves in **Fig. 15(a)** and **Fig. 15(c)**, it can be seen that, the WBCE converges faster and more stable than the BCE, and moreover, the smoother curves also prove that WBCE can effectively overcome the problem of gradient vanishing/explosion at the initial stage of training.

From the accuracy curves of **Fig. 15(b)** and **Fig. 15(d)**, it can be seen that, the BCE curve varies greatly in the early stage of training, which indicates an unstable training. On the contrary, the WBCE performs smoother and more stable.

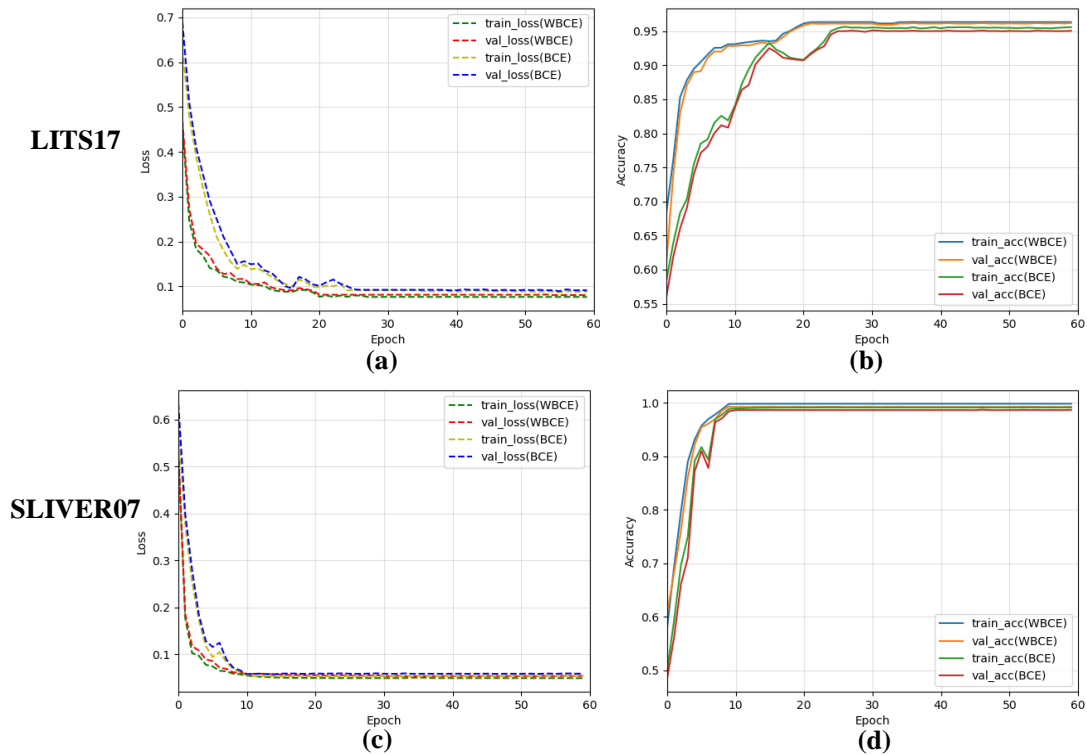

**Fig. 15.** Loss and accuracy curves of training and validation using WBCE and BCE. (a) Loss value on LiTS17 (b) Accuracy on LiTS17 (c) Loss value on SLiver07 (d) Accuracy on SLiver07

In addition, **Fig. 16** shows some typical segmentation results of SAR-U-NET using BCE and WBCE, respectively. It can be seen that, the WBCE enables more accurate segmentation results. The main reason can be attributed to the employment of the weight factor, which makes the model pay more attention to the liver segmentation, and effectively avoid over/under-segmentation.

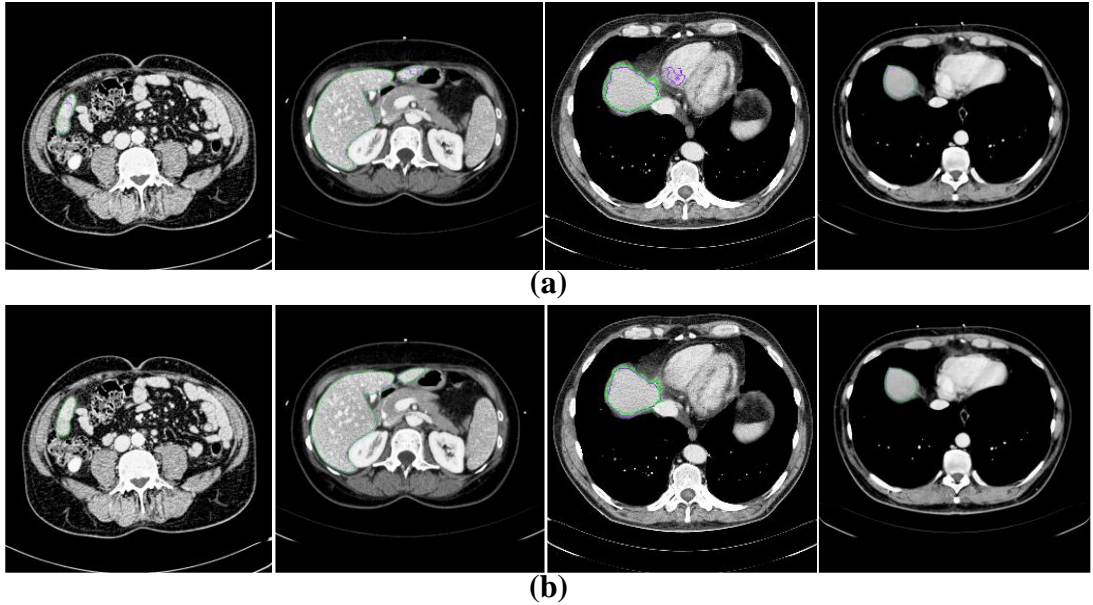

**Fig. 16.** Comparative segmentation results with/without wight factor. (a) without wight factor, BCE (b) with wight factor, WBCE

### 4.8 Comparison of running time of different models

**Table 7** and Table 8 list the total training and testing time (per case) of different models on datasets of LiTS17 and SLiver07, respectively. It can be seen from the tables that, the training and testing time of the proposed SAR-U-Net are the highest. Specifically, due to the introductions of residual module, SE block and ASPP in the ablation experiment, the training and testing time of SAR-U-NET is gradually increased compared with the other methods. However, more image features are obtained through deeper network, and the accuracy of the proposed SAR-U-Net is greatly improved, which has been verified in Section 4.5. On the whole, the training time of SAR-U-NET is higher than other models, while the test time is only slightly higher than other models. Nevertheless, this strategy of improving accuracy at the cost of time is still relatively meaningful in computer-aided diagnosis.

**Table 7.** Training and testing time (per case) of different models on LiTS17 dataset

| Networks | Training time | Testing time (per case) |
|---|---|---|
| FCN | 31h 48min | 34.5sec |
| U-Net | 38h 26min | 45.2sec |
| Attention U-Net | 49h 32min | 1min 8.5sec |
| ResU-Net | 53h 25min | 1min 21.8sec |
| SE-ResUNet | 59h 18min | 1min 37.3sec |
| ASPP-ResUnet | 64h 39min | 1min 56.6sec |
| SAR-U-Net | 71h 46min | 2min 7.2sec |

**Table 8.** Training and testing time (per case) of different models on Sliver07 dataset

| Networks | Training time | Testing time (per case) |
|---|---|---|
| FCN | 26h 15min | 25.6sec |
| U-Net | 30h 50min | 36.6sec |
| Attention U-Net | 43h 23min | 53.3sec |
| SAR-U-Net | 51h 56min | 1min 8.6sec |

### 4.9 Limitations

Although encouraging results were obtained by our proposed SAR-U-Net framework, limitations still exist for further enhancement. Our proposed method is based on 2D network, but the CT images are 3D-based, thus it is prone to miss important context information on the z-axis. In addition, once the liver edge contains lesions or tumor abnormalities (as shown in **Fig. 17**), the proposed method may result in large errors around the boundary. Regarding such limitations, evolving the current 2D framework to 3D would further improve the accuracy via context of z-axis, and thus reduce the errors around the liver abnormalities.

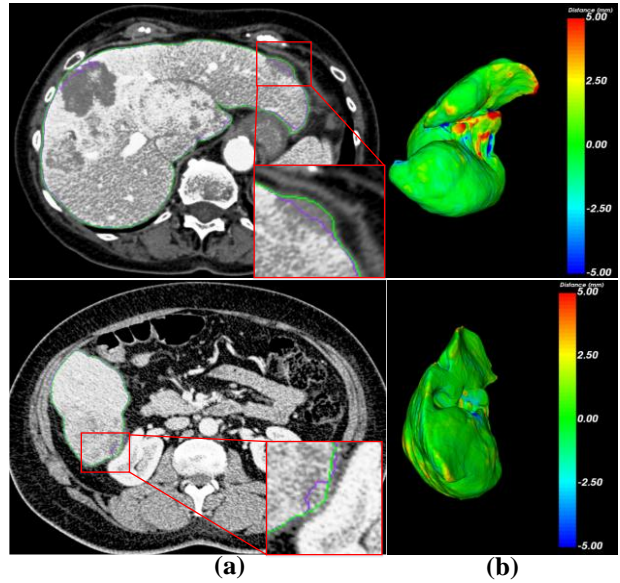

**Fig. 17.** Limitations on liver boundary lesions of proposed method. (a) 2D error visualization (the green line indicates the gold standard, while the purple line indicates our results) (b) 3D error visualization (Blue and red indicate large over-segmentation and under-segmentation errors, while green indicates small errors.)

## 5 Conclusion

This paper proposes a novel SAR-U-Net network for automatic liver segmentation from CT. The U-Net based framework leverages the advantages of Squeeze-and-Excitation, Atrous Spatial Pyramid Pooling, and residual learning techniques. In order to obtain more important image features, we add SE module to each convolution unit of U-Net coder to make the network learn image features adaptively and suppress irrelevant areas, so that the network pays more attention to the related features of segmentation task. In addition, to increase the receptive field of the network and learn multi-scale features, the ASPP module is employed as the transition layer and output layer of the network to achieve multi-scale feature extraction. Furthermore, the residual structure is applied to the network to enable a deeper network and to avoid the possible gradient vanishment problem, and the WBCE loss function is employed for a more targeted and effective training.

The high accuracy and robustness of our proposed model were verified on two competition training datasets, LiTS17 and SLiver07. Compared with the related classic methods, the proposed method achieved superior performance on the quantitative metrics. Moreover, compared with other models, the proposed SAR-U-NET method showed the smoothest curves on accuracy and loss, with the fastest convergence speed. Through the ablation experiment, the proposed SAR-U-NET obtained the optimal metrics, which proved the effectiveness of the improved strategy. Besides, the effectiveness of the weighted cross-entropy loss function was verified, which encouraged the network to pay more attention to the target area, thereby effectively avoiding the occurrence of over-/under- segmentation. In addition, in terms of the training and testing time, although the proposed SAR-U-Net was higher than other methods, the overall performance was significantly improved. Specifically, the proposed method showed a significant improvement and robustness in handling small liver regions, discontinuous liver regions and blurry liver boundaries. Nevertheless, our proposed method is prone to slight over-segmentation or under-segmentation errors, when dealing with lesions or tumors around liver boundary, and thus make full use of the z-axis information in 3D to reduce errors would be the focus of our future work.